%% file: main.tex
\documentclass{llncs}
\usepackage[T1]{fontenc}
\usepackage{anyfontsize}
\usepackage{graphicx}
\usepackage[bookmarksopen]{hyperref}

\usepackage{cite}
\usepackage{amsmath,amssymb,amsfonts}
\usepackage{algorithmic}
\usepackage{textcomp}
\usepackage{xcolor}
\usepackage{balance}

\usepackage{caption}
\usepackage{subcaption}
\usepackage{booktabs}

\usepackage{mathabx}

\usepackage{listings}

\definecolor{codegreen}{rgb}{0,0.6,0}
\definecolor{codegray}{rgb}{0.5,0.5,0.5}
\definecolor{codepurple}{rgb}{0.58,0,0.82}
\definecolor{codeblue}{rgb}{0.1,0.3,0.7}
\definecolor{backcolor}{rgb}{0.95,0.95,0.92}
\definecolor{formalfill}{HTML}{FFF3DC}
\definecolor{formaldraw}{HTML}{B07A1E}
\definecolor{testfill}{HTML}{EDEDED}
\definecolor{testdraw}{HTML}{7A7A7A}

\lstdefinelanguage{Rust}{
  morekeywords={as, async, await, break, const, continue, crate, dyn, else, enum,
    extern, false, fn, for, if, impl, in, let, loop, match, mod, move, mut, pub, ref,
    return, self, Self, static, struct, super, trait, true, type, unsafe, use, where,
  while},
  morekeywords=[2]{bool, char, str, u8, u16, u32, u64, u128, usize, i8, i16, i32, i64,
  i128, isize, f32, f64, String, Vec, Option, Result, Box, Some, None, Ok, Err},
  sensitive=true,
  morecomment=[l]{//},
  morecomment=[n]{/*}{*/},
  morestring=[b]",
}

\lstdefinestyle{rusty}{
  backgroundcolor=\color{backcolor},
  commentstyle=\color{codegreen},
  keywordstyle=\color{codeblue},
  numberstyle=\tiny\color{codegray},
  stringstyle=\color{codepurple},
  basicstyle=\ttfamily\fontsize{7.5}{8.5}\selectfont,
  breakatwhitespace=false,
  breaklines=true,
  captionpos=b,
  keepspaces=true,
  numbers=left,
  numbersep=5pt,
  showspaces=false,
  showstringspaces=false,
  showtabs=false,
  tabsize=2,
  upquote=true,
  language=Rust,
  keywordstyle=[2]\color{codepurple},
}

\usepackage{tikz}
\usetikzlibrary{calc,positioning,arrows,fit,patterns,backgrounds,
shapes.misc,shapes.symbols,arrows.meta}

\definecolor{trusted}{HTML}{33FF33}
\definecolor{groupfill}{HTML}{F7F3EA}
\definecolor{groupdraw}{HTML}{B9A88A}
\definecolor{grouptext}{HTML}{56708C}
\definecolor{toolfill}{HTML}{E9E4F5}
\definecolor{tooldraw}{HTML}{8F82B8}
\definecolor{traitfill}{HTML}{f6f78b}
\definecolor{userfill}{HTML}{FBE4E8}
\definecolor{userdraw}{HTML}{C2687D}

\tikzset{
  type/.style={rectangle, draw, fill=blue!20,
  font=\footnotesize, inner sep=3pt, minimum height=5mm},
  trait/.style={chamfered rectangle, chamfered rectangle xsep=2mm,
    draw, fill=traitfill, font=\footnotesize\itshape,
  inner sep=2.5pt, minimum height=5mm},
  user/.style={rectangle, rounded corners=2.5mm, draw=userdraw, dash pattern=on 4pt off 1.5pt, line width=0.9pt,
  fill=userfill, font=\footnotesize, inner sep=3pt, minimum height=5mm},
  datatype/.style={rectangle, draw, fill=white, font=\footnotesize\ttfamily,
  inner sep=2pt},
  impl/.style={-{Stealth[open, length=2mm]}, dashed},
  extends/.style={-{Triangle[open, length=2mm, width=2mm]}},
  call/.style={-{Stealth[length=2mm]}},
  pass/.style={-{Stealth[length=2mm]}, dotted, thick},
  rel/.style={font=\scriptsize, midway, auto, inner sep=1.5pt},
  group/.style={rectangle, draw=groupdraw, line width=0.6pt,
  fill=groupfill, inner sep=2.5mm},
  group label/.style={font=\footnotesize, text=black, inner sep=1.5pt,
  anchor=south west},
  env/.style={group},
  env label/.style={group label},
  trusted region/.style={draw, thick, solid, pattern=north west lines,
  pattern color=trusted, inner sep=2mm},
  trusted label/.style={font=\scriptsize, anchor=south},
  component/.style={type},
  tool/.style={rectangle, draw=tooldraw, fill=toolfill, font=\footnotesize\scshape,
  inner sep=3pt, minimum height=5mm},
  tool optional/.style={tool, densely dotted, fill=toolfill!45, text=black!70},
  test/.style={draw=testdraw, fill=testfill, font=\footnotesize,
  inner sep=3pt, minimum height=6mm},
  formalisation/.style={rectangle, draw=formaldraw, double=formalfill,
    double distance=1.2pt, fill=formalfill, font=\footnotesize,
  inner sep=3.5pt, minimum height=5mm},
  proved/.style={-{Implies}, double, double distance=1.2pt, shorten >=1pt},
  tested/.style={-{Stealth[length=2mm]}, dash dot},
  equal/.style={{Implies}-{Implies}, double, double distance=1.2pt,
  shorten >=1pt, shorten <=1pt},
}

\newcommand{\smt}[1]{`\texttt{#1}'}

\usepackage{comment}

\usepackage{multirow}
\usepackage{float}
\usepackage{orcidlink}

\usepackage{pseudocode}
\usepackage{todonotes}
\setuptodonotes{
  color=gray!12,        
  bordercolor=gray!60,  
  linecolor=gray!60,    
  textcolor=black!80,
  size=\footnotesize,
}

\makeatletter\@twosidefalse\@mparswitchfalse\makeatother
\begin{document}
\title{Blazing-Fast SMT-LIB Parsing for Rust and Lean}
\author{Jan Onderka\inst{1}\orcidlink{0000-0003-2069-8584} \and
  Armin Biere\inst{1}\orcidlink{0000-0001-7170-9242} \and
Henrik Böving\inst{2}\orcidlink{0009-0007-8294-2710}}
\institute{University of Freiburg, Germany\\
  \email{\{onderka,biere\}@cs.uni-freiburg.de}
  \and Lean FRO, Germany\\
\email{henrik@lean-fro.org}}
\maketitle

\begin{abstract}
  SMT-LIB is a standard input format
  for Satisfiability Modulo Theories (SMT) solvers.
  However, parsing the format is non-trivial
  and can become a bottleneck,
  e.g. when retrieving benchmark metadata or
  normalising the input for solvers.
  We introduce blazing-smtlib, a Rust parsing library for SMT-LIB,
  and BSmtLib, its port to Lean.
  We outline the library API we designed for composability
  and extensibility (including non-standard SMT-LIB commands),
  and discuss our implementation decisions for fast lexing and parsing.
  We proved that our lexing and parsing with Lean-specific optimisations
  match our formalised reading of the SMT-LIB syntax.
  We measured performance on a representative sample from
  the 2025~SMT-LIB corpus and found that blazing-smtlib
  is more than 6 times faster on the geometric mean
  than previous parsing libraries
  and BSmtLib is faster or comparable to them.
\end{abstract}


\input{sections/01_introduction}
\input{sections/02_background}
\input{sections/03_api}

\input{sections/04_rust}

\input{sections/05_lean}

\input{sections/06_evaluation}
\input{sections/07_conclusion}

\balance
\bibliographystyle{splncs04}
\bibliography{main}

\clearpage
\appendix
\input{sections/a_corpus}
\input{sections/b_measurements}

\end{document}

%% file: sections/01_introduction.tex
\section{Introduction}
\label{sec:introduction}

Satisfiability Modulo Theories (SMT) solving is an important and rich topic in verification
and automated reasoning.
The almost universal input format for SMT solvers is SMT-LIB~2 \cite{smtlib2},
with a wide corpus of SMT benchmarks \cite{smtlib-nonincremental,smtlib-incremental}
that can be used to test the solvers
and is used to compare their performance in the annual SMT solver competition (SMT-COMP) \cite{smt-comp}.

While SMT-LIB~2 is human-readable and it is fairly easy to implement a parser
that reads the standard benchmarks, writing a specification-compliant parser
that covers all syntactic constructs is hard.
There are SMT-LIB parsing libraries available for some programming languages
but they tend to lack extensibility for non-standard commands
that solvers with custom parsers can support and
are not thoroughly optimised for performance.

We present \textbf{blazing-smtlib}, a high-performance SMT-LIB parsing library written in the
programming language Rust, and its port \textbf{BSmtLib}
in the Interactive Theorem Prover (ITP) language Lean~\cite{lean4}.

Both libraries are composable and extensible,
allowing implementation of solvers with non-standard commands,
reparsers, pretty-printers, response parsers etc.
They are zero-dependency (only using the standard library of Rust / Lean),
contain no unsafe code, and
are tested on
a variety of valid and invalid inputs and
the 2025~SMT-LIB corpus.
\textbf{BSmtLib} lexing and parsing
is additionally proven to match our reading
of the SMT-LIB syntax formalised in Lean.
On a geometric mean over a representative sample
from the 2025~SMT-LIB corpus,
\textbf{blazing-smtlib} is more than 6 times faster
than previous parsing libraries
and \textbf{BSmtLib} is faster or comparable to them.

%% file: sections/02_background.tex
\section{Background and Related Work}
\label{sec:background}

As of the time of the writing, there are four minor versions
of SMT-LIB~2~\cite{smtlib2} that change various details,
but keep the basic shape and mostly preserve backwards compatibility:
2.0~\cite{smtlib20}, 2.5~\cite{smtlib25}, 2.6~\cite{smtlib26}, and 2.7~\cite{smtlib27}.
Each version may have multiple revisions to fix bugs.
The specification of each revision is given in one PDF file.
In the rest of this paper, we will refer to SMT-LIB~2 as ``the standard''
and reference only the minor versions and revisions when necessary.

While the SMT-LIB~2 format is human-readable and can be written fairly easily by hand or a tool,
parsing it is a challenge.
When introducing the solver \textbf{cvc5}, the authors recognised the need for a faster parser
to replace the one inherited from \textbf{CVC4}~\cite[sec.~5]{barbosa2022}.
In recent work on SMT-LIB input normalisation,
the used parser timed out in 1 minute on 58 benchmarks~\cite[sec.~5]{amrollahi2025}.
A database containing SMT-LIB corpus metadata
from benchmark files has been created~\cite{schurr2026},
the metadata retrieval described as the most time-consuming stage~\cite[sec.~4]{schurr2026}.

It is common to run parallel solver portfolios~\cite{schreiber2026}.
A portfolio of same-solver instances
can run parsing once~\cite{wintersteiger2009}
or multiple times~\cite{barbosa2022}.
Using multiple different solvers in parallel~\cite{backes2018}
or selecting the solver based on the input~\cite{scott2021,pimpalkhare2021}
requires multiple parsing passes (for each solver or the selector and solver)
when using SMT-LIB scripts.
In multi-solver tools, the I/O cost of the parsing led to the creation~\cite{pereira2026}
and use~\cite{pereira2026, palikareva2013}
of frontend APIs that bypass the format.

Many SMT-LIB parsing libraries exist for a variety of programming
languages~\cite{bonichon2015,gario2015,smtlib2parser,smtlib-utils,smt2parser,yaspar,language-smtlib}.
Notably, \textbf{Dolmen}~\cite{bury2021} allows lexing, parsing, and well-formedness checking
of SMT-LIB scripts. It is written in OCaml, considered the reference parser
and type-checker for SMT-LIB~\cite{schurr2026}, and used
as the model validator for the SMT solver competition~\cite{smt-comp-rules}.

Mainline SMT solvers in SMT-COMP,
including \textbf{cvc5}~\cite{barbosa2022}, \textbf{Z3}~\cite{moura2008},
\textbf{Yices}~\cite{dutertre2014},
and \textbf{Bitwuzla}~\cite{niemetz2023}, all employ custom parsers.
Custom parsers can be extended to support non-standard commands such as
\smt{minimize} and \smt{maximize} provided by Z3~\cite{bjorner2015}
or those added in SV-LIB,
a standard exchange format for software verification that extends SMT-LIB~2.7~\cite{beyer2025}.
Unfortunately, extensibility of parsing libraries to non-standard commands has been limited:
\textbf{Dolmen} and \textbf{language-smtlib}~\cite{language-smtlib}
can only process non-standard arguments as untyped expressions,
\textbf{pySMT}~\cite{gario2015} can only be extended by an internal table,
and most other libraries cannot be extended at all.

Another area of potential use of SMT-LIB is within Interactive Theorem Provers (ITPs),
where an SMT solver is used as a \emph{hammer}~\cite{meng2006,blanchette2011}
and the result of its run on an SMT-LIB script is verified against a trusted specification.
SMT-LIB benchmarks could for instance also be used
during development of new ITP tactics for testing.
Cursory unverified parsers have been implemented in Lean as a part
of \textbf{lean-auto}~\cite{yicheng2025},
\textbf{lean-smt}~\cite{abdalrhman2025}
and the \textbf{Leanwuzla}~\cite{leanwuzla} entry point for the
\textbf{bv\_decide} tactic~\cite{boving2025}.
Major hurdles along this line of work include the lack of an official SMT-LIB formalisation
and the fact that ITP programs can easily be an order of magnitude slower than their counterparts
in languages such as C, C++, or Rust, especially in low-level code such as parsing.

This work was motivated by us encountering
the aforementioned problems while developing our twin tools \textbf{Roole}
(an unverified solver written in Rust)
and \textbf{Roolean} (a verified proof checker written in Lean)~\cite{onderka2026twin,roole,roolean}.
Due to the lack of Lean libraries for SMT-LIB parsing, we implemented an SMT-LIB parser of our own.
In \textbf{Roole} 0.1.1, we switched the used parser from \textbf{aws-smt-ir}~\cite{aws-smt-ir}
to a port of the \textbf{Roolean} parser.
Witnessing the value of a fast, extensible, and composable parsing library,
we started development of \textbf{blazing-smtlib}
from the \textbf{Roole} 0.1.1 parser.
We then ported \textbf{blazing-smtlib} back to Lean under the name \textbf{BSmtLib}
and optimised it specifically for Lean, minimising the potential for
optimisation-related bugs through testing and proving it correct against
our formalised reading of SMT-LIB syntax.

\subsubsection{Overview.}
As \textbf{blazing-smtlib} and \textbf{BSmtLib} are based on the same concepts
with language-specific differences,
we discuss the API of \textbf{blazing-smtlib} in Section~\ref{sec:api},
its speed-optimised implementation in Section~\ref{sec:rust},
and the \textbf{BSmtLib} port in Section~\ref{sec:lean}.
Finally, we evaluate their performance in Section~\ref{sec:evaluation}.

%% file: sections/03_api.tex
\begin{figure}[tb]
  \centering
  \begin{tikzpicture}[node distance=0.7cm and 0.7cm]

    \node[user] (driver) {Driver};
    \node[type] (lexer) [below = of driver]{Lexer};
    \node[user] (token-provider) [right = of lexer]{Token provider};

    \node[trait]  (token-consumer) [below = of lexer] {Consume};

    \node[type] (parse-adapter) [below = of token-consumer] {ParseAdapter};
    \node[user]  (user-consumer) [right = of parse-adapter] {Token consumer};

    \node[trait] (parse) [below = of parse-adapter] {Parse};

    \node[type] (command-parser) [below left = 0.7cm and -0.5cm of parse] {CommandParser};
    \node[type] (term-parser) [right = of command-parser] {TermParser};
    \node[type] (script-parser) [left = of command-parser] {ScriptParser};
    \node[type] (other-parser) [right = of term-parser] {...};
    \node[user] (user-parser) [right = of other-parser] {Custom parser};

    \node[trait] (handler) [below = of other-parser] {Handler};
    \node[trait] (command-handler) [below = of command-parser] {CommandHandler};

    \node[user] (user-handler) [below = 2cm of term-parser] {User handler};


    \draw[call] (driver) -- node[rel] {lex} (lexer);

    \draw[call] (lexer) -- node[rel] {consume} (token-consumer);
    \draw[call] (token-provider) -- node[rel] {consume} (token-consumer);

    \draw[impl] (token-consumer) -- (parse-adapter);
    \draw[impl] (token-consumer) -- (user-consumer);

    \draw[call] (parse-adapter) -- node[rel] {parse} (parse);

    \draw[impl] (parse) -- (script-parser);
    \draw[impl] (parse) -- (command-parser);
    \draw[impl] (parse) -- (term-parser);
    \draw[impl] (parse) -- (other-parser);
    \draw[impl] (parse) -- (user-parser);

    \draw[extends] (command-handler) -- (handler);

    \draw[call] (command-parser) -- node[rel] {handle} (command-handler);
    \draw[call] (script-parser) -- node[rel] {handle} (command-handler);
    \draw[call] (term-parser) -- node[rel] {handle} (handler);
    \draw[call] (other-parser) -- node[rel] {handle} (handler);
    \draw[call] (user-parser) -- node[rel] {handle} (handler);

    \draw[impl] (handler) -- (user-handler);
    \draw[impl] (command-handler) -- (user-handler);

    \node[font=\scriptsize, align=left, anchor=north west]
    at (script-parser.west |- driver.north) {%
      \tikz\draw[call] (0,0) -- (1cm,0); calls\\
      \tikz\draw[impl] (0,0) -- (1cm,0); implemented by\\
      \tikz\draw[extends] (0,0) -- (1cm,0); extends\\
      \tikz\node[user, rounded corners=1mm, minimum height=3mm, inner sep=1pt] {\phantom{x}}; user-implemented
    };

    \begin{scope}[on background layer]
      \node[group, fit=(lexer) (token-consumer) (parse-adapter) (parse)] (reader) {};
      \node[group label] at (reader.north west) {Reader};
    \end{scope}
  \end{tikzpicture}
  \caption{
    Simplified dataflow of the building blocks and user entry points in \textbf{blazing-smtlib},
    without the \texttt{Smt} prefix for better readability.
    The user can provide
    the input as a file or stream of bytes (driver) or as a stream of tokens (token provider).
    The user can implement interfaces to
    process tokens (token consumer), parse tokens (custom parser), and handle parsed
    SMT-LIB categories and commands (user handler). A custom parser can be composed
    of provided category parsers and combinators, and the user handler can select
    a parser for parsing non-standard commands.
  }
  \label{fig:api}
\end{figure}
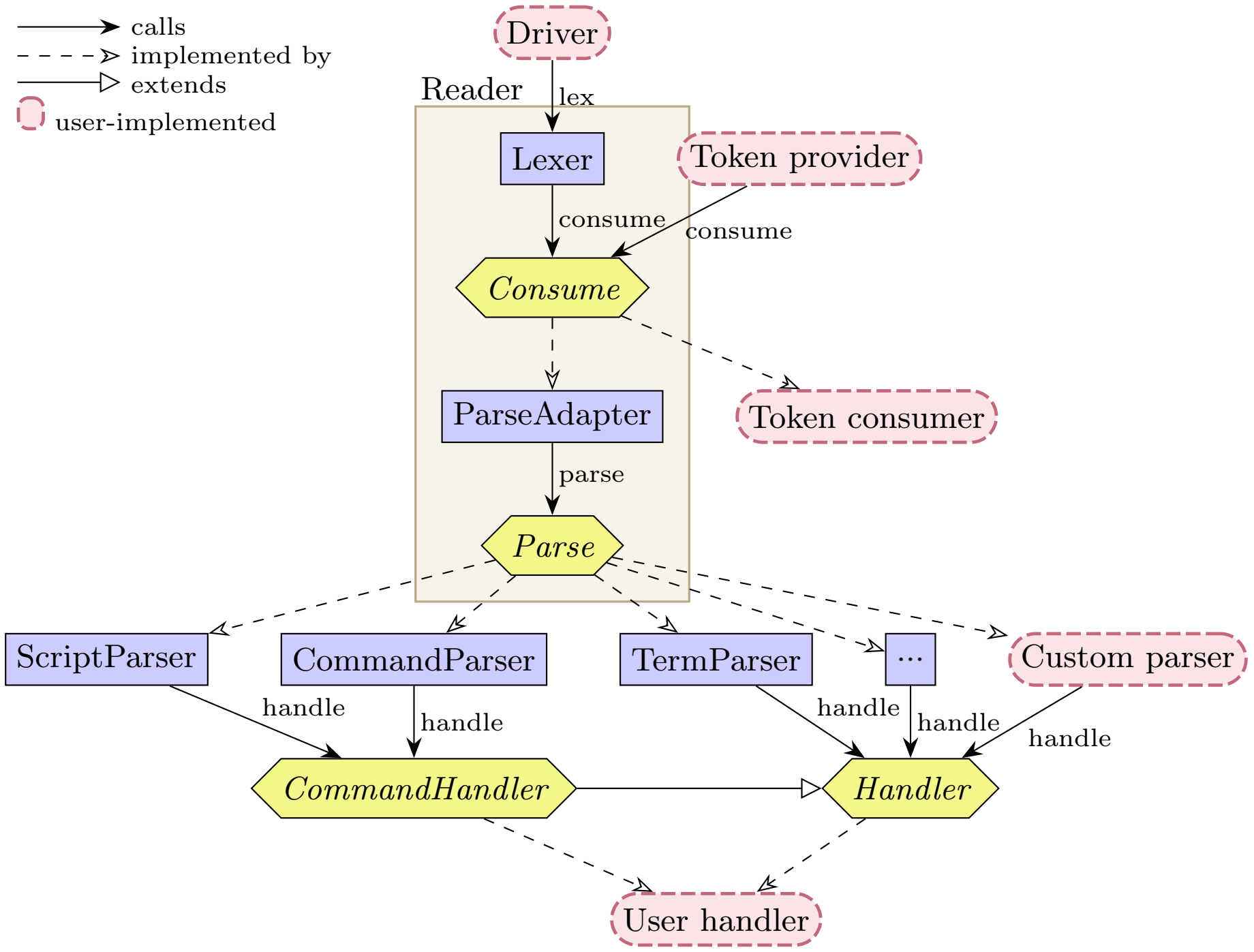

\section{Library Interface}
\label{sec:api}

There are two main types of parsing libraries:
those that build a syntax tree, usually an abstract syntax tree (AST),
and those that run over the stream of bytes and call a given handler
on each syntax node (event-based API).
The former are a good fit for languages with complicated parsing and small files,
the latter for languages with simpler parsing but large files.
SMT-LIB~2 fits the latter: it has Lisp-like syntax
and the largest file in the corpus is approx.~2 GB.
AST-building libraries are forced to retain the whole tree in memory.

Our library \textbf{blazing-smtlib} is a push-style streaming library
in the style of Simple API for XML~(SAX)~\cite{sax},
with focus on speed and composability.
A properly composable library is difficult to implement but can fulfil the needs of more users,
and testing and reasoning (e.g. when proving Lean theorems) are simplified due to the decreased coupling.

Figure~\ref{fig:api} shows the basic types and interfaces provided by \textbf{blazing-smtlib}.
Our library is push-style (data flows into the later stages), so the arrows describe the dataflow:
a driver hands data to \texttt{SmtLexer}, which calls an instance of a \texttt{SmtConsume}.
This instance might be e.g. \texttt{SmtParseAdapter}, which calls an instance of a \texttt{SmtParse}, and so on.
This gives the user five entry points into the library instead of the usual two
(bytestream/file input and parsed categories/commands)
or three (token consumption, if the lexer can be used separately).
In addition to e.g. token-category counting or token-level pretty-printing
usable through a separate lexer, the user can write their custom
token provider (e.g. translating a problem written in another language
  without the need to go to text level
or lexing a custom SMT-LIB-like binary format),
and add new parsers, composable out of standard ones
(for processing non-standard commands,
reading non-script files such as solver responses, etc.).

Still, the typical usage scenario is writing an SMT-LIB solver. This can be done in a few simple steps.
First, the user implements \texttt{SmtHandler}, setting the handler error type, the types that will be returned
after handling, a binder scope type,
and functions to handle each occurrence of a syntactic category:

\noindent
\begin{minipage}{\linewidth}
\begin{lstlisting}
impl SmtHandler for UserHandler {
  type Error = UserError;

  type SExpr = UserSExpr;
  type Attribute = UserAttribute;
  type Sort = UserSort;
  type Term = UserTerm;

  type Scope = UserScope;

  fn set_version(&mut self, version: SmtVersion) {
      /* implement */
  }

  fn handle_sort(&mut self, sort: SmtSort<'_, Self>)
    -> Result<UserSort, UserError> {
      /* implement */
  }
  fn handle_term(&mut self, term: SmtTerm<'_, Self>)
    -> Result<UserTerm, UserError> {
      /* implement */
  }

  fn start_let(&mut self,
    bindings: SmtBindings<'_, UserTerm>)
    -> Result<UserScope, UserError> {
      /* implement */
  }
  fn end_let(&mut self,
    scope: UserScope, body: UserTerm)
    -> Result<UserTerm, UserError> {
      /* implement */
  }
  // handlers for other syntactic categories
  (...)
}
\end{lstlisting}
\end{minipage}

The parsers send events corresponding to post-order AST traversal to the handler,
meaning that parsing \smt{(add a b)} will result in a call to \texttt{handle\_term} with the term parameter
containing the symbol \smt{a}, then in a call with \smt{b}.
Typically, the user will find and return the unique identifiers $a$ and $b$ (values of \texttt{UserTerm})
that correspond to the symbols in a map of definitions.
After all of the parameters are processed, \texttt{handle\_term} will be called for
\smt{(add a b)} itself, with the function name \texttt{add}
and a parameter array composed of $a$ and $b$.

Let expressions, quantifiers and lambdas are slightly more involved:
given \smt{(let ((x a) (y b)) x)}, the term parser first collects
the binding term values $a$ and $b$ by calling \texttt{handle\_term},
then lets the user introduce a let scope by calling \texttt{start\_let}
with the collected bindings [(\smt{x}, $a$), (\smt{y}, $b$)].
The user can use the custom \texttt{UserScope} type to store information
that will be used when closing the scope
using \texttt{end\_let}. Using a map of definitions,
the user can replace the previous mapping, store it in \texttt{UserScope},
and restore it at \texttt{end\_let}
without needing to keep a custom stack of scopes.
The user can also use a zero-sized scope type,
avoiding any performance penalty
when managing scopes manually.

After implementing the category handler, the user implements a command handler,
deciding what should be done when commands are issued:

\noindent
\begin{minipage}{\linewidth}
\begin{lstlisting}
impl SmtCommandHandler for UserHandler {
  type NonstandardParser = UserParser;

  fn command_assert(&mut self, term: UserTerm)
    -> Result<(), UserError> {
      /* implement */
  }
  // handlers for other commands
  (...)

  fn nonstandard_command(
    &mut self,
    name: SmtSymbol<'_>,
  ) -> Result<UserParser, UserError> {
    Ok(UserParser::new())
  }
}
\end{lstlisting}
\end{minipage}

The standard commands can be implemented similarly to categories,
with e.g. \texttt{command\_assert} adding the given term to current assertions.
On success, they return no data, as they are used to change the current semantic
state carried in the handler or produce other effects, e.g. print some output.
A handler implementation that does not interact with the global state
(e.g. with output streams adjustable on construction) could also be used through
the API alone by tools such as portfolio runners, as the types such as
\texttt{SmtTerm} or \texttt{SmtBindings} can be constructed outside of our library.

The special \texttt{nonstandard\_command} method is called when a command
with a non-standard name is encountered.
The user can either reject the command or return a parser that will parse the rest of it.
The parser can be a standard parser, a composition of standard parsers
or a user-written parser that can process tokens specially and/or call standard parsers itself.
As standard category parsers return a properly typed value once they
have finished processing it
(e.g. \texttt{SmtTermParser} returns a \texttt{UserTerm}),
it is easy to extend a solver with extensions such as
\smt{(maximize t)}, where \smt{t} stands for an arbitrary term.

While the composability can be used for maximum control,
we care about simple usage in usual scenarios.
For the caller, we provide \texttt{SmtReader} that directly connects
the driver with the handler,
combining lexing and parsing.
A step further, reading SMT-LIB files can be as easy as
implementing a command handler and writing
\begin{lstlisting}
read_smt_file(&mut user_handler, file_path)
\end{lstlisting}
which reads the SMT-LIB file and calls the handler with the appropriate categories and commands to handle.

\subsubsection{Whitespace.}
We provide an option for the token consumers to request whitespace and comments to be emitted as tokens,
which allows printing while preserving some parts of the layout, up to 1:1 reconstruction of input files\footnote{
  With a single exception of SMT-LIB~2.0 files containing occurrences of non-escaped backslashes in strings,
  which tokenise to the same content as properly escaped backslashes.
  Files for SMT-LIB versions 2.5+ can be reconstructed exactly.
}.

\subsubsection{Continuation and errors.}
Our parsers and parser adapter are designed to reset on producing a result,
ready to process another piece of the same data (self-reset).
On encountering an error, we implement fail-fast behaviour,
where processing is stopped in an unspecified state
and continuing to process is not allowed.
We have done a preliminary exploration of error recovery
allowed by the standard, but it leads to ambiguity.
For example, in \smt{|x\textbackslash}, it is unclear if the disallowed backslash was written
instead of the closing pipe of the quoted symbol by mistake, or if it is spurious and the quoted token
should continue. As such, we currently do not provide an option for recovery.
For the usual case of generated SMT-LIB files, an error is a sign that something is wrong with
the generator or the data stream, and error recovery should not be performed.

\subsubsection{Locations.}
The token provider hands over cursors over tokens that implement a token interface to token consumers,
and can add additional information that can be used non-generically. Our lexer adds offset/line/column
token span, and returns or wraps errors with the span of the token where the error occurred.
Generic parsers do not interact with location or other information in tokens, and it is not passed to
the handlers to simplify signatures and increase speed.

A parser adapter specific to our lexer could be used to use the location information more extensively,
for example by tracking parentheses and delaying an error to provide the full
command or term location where the error occurred.
We did not add such constructions to the library to keep it simple and avoid adding code
that would not be used by most users.

%% file: sections/04_rust.tex
\section{Rust Implementation}
\label{sec:rust}

The SMT-LIB~2 language is a Lisp-like language with a separate lexicon over 8-bit bytes (US-ASCII with the encoding of higher bytes unspecified)
and a grammar over the lexical tokens, read in a streaming fashion in a Read-Eval-Print loop (REPL).
It is unusual in that it can be considered a programming language (semantically expressing program or system behaviour)
or a data interchange language (as the major use is running automatic tools that receive automatically generated files).
For lexer or parser construction, it is more useful to consider it a data interchange format and apply standard techniques.
Our lexer and parsers are fairly standard constructions with two
crucial speed-improving techniques:
\begin{itemize}
  \item \textbf{Batch processing.}
    The token consumers / parsers receive a cursor from which they consume tokens.
    The lexer is given a slice of bytes in one call,
    creates the tokens and puts them into a batch,
    then calls the consumer with a cursor into the batch,
    minimising the token handover work.
    Batch processing is intentionally not used for handlers for a cleaner,
    easy-to-use API, with the potential speed benefits lower
    in the later stage.

  \item \textbf{Zero-copy dispatch.}
    When possible, the lexer produces token batches that only reference data
    (symbol names, literal values, etc.) through a slice into the input byte buffer without any copying,
    and highly-used parsers do not copy when it is possible to call the handler with the reference.
\end{itemize}
\begin{figure}
  \includegraphics{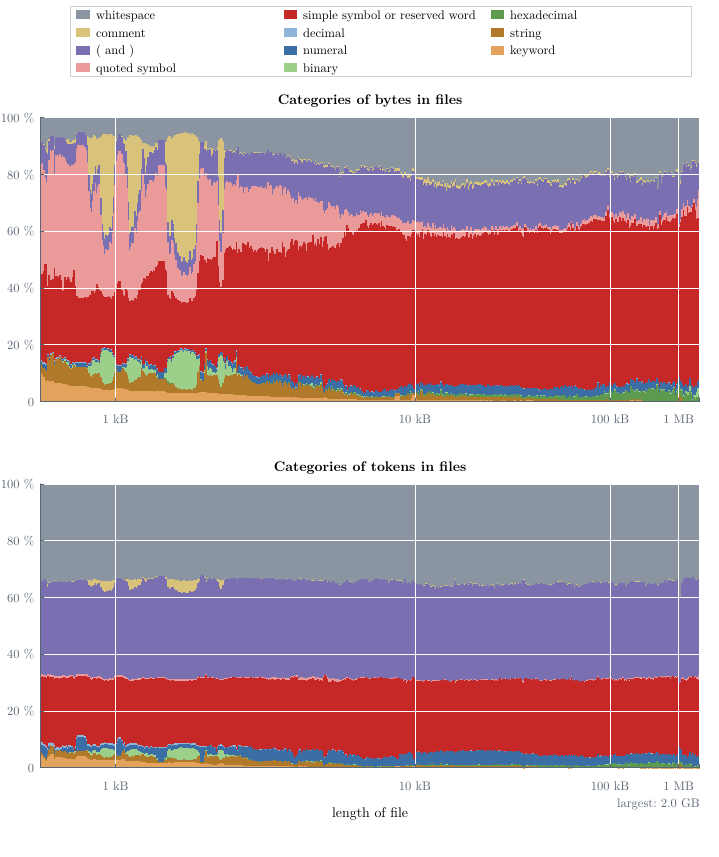}
  \caption{Shares of bytes and tokens per category on 2025~SMT-LIB corpus
    files,
    as seen by our lexer on its input and its dispatch arms.
    Reserved words are lexed the same
    as simple symbols, with a reserved-word lookup afterwards; we move
    the lookup to the parser side for easier reconstruction.
    Even if whitespace and comments are not emitted, they must be processed by the lexer.
    Whitespace, parentheses, and symbols dominate and must be processed quickly.
    Long comments comprise a large part of some small files.
    The other categories are less relevant for lexer speed.
    The shares of token categories are regular due to the S-expression shape,
    e.g. a 2-argument function application \smt{(bvadd a b)}
    consists of 2 parentheses, 3 simple symbols, and 2 whitespaces.
  }
  \label{fig:file-shares}
\end{figure}

\subsection{Lexing}
\label{subsec:lexing}

Our lexer is a simple iterative state-machine lexer with additional machinery.
The speed is achieved by a single eager zero-copy dispatch loop in the normal lexer state.
When the token content is a slice of the input buffer,
the token is added to the current batch of tokens and the loop continues.
The loop is terminated and the batch handed over to the consumer only when
adding another token becomes impossible
(e.g. a string contains an escape or lexing needs to continue past the buffer end).
In practice, over 99.99\% of tokens are produced in the loop, so it is the only place
requiring heavy optimisation. Due to the unequal shares of lexical categories
shown in Figure \ref{fig:file-shares}, they are tested in a preferential order:
\begin{enumerate}
  \item closing parenthesis,
  \item opening parenthesis,
  \item symbol (simple, reserved word, quoted) or whitespace,
  \item everything else.
\end{enumerate}

The closing parenthesis is tested first as there are long runs (successive occurrences)
of closing parentheses, compared to opening parentheses where there are no non-trivial
runs \footnote{Appendix \ref{app:corpus} contains additional figures for token prevalence, length, and runs.}.
A long run of closing parentheses is then processed as fast as possible,
running through the fastest path with accurate branch prediction.

Non-parenthesis tokens can be arbitrarily large, so it is necessary to
move after the run of bytes allowed in them as fast as possible for
good byte-dispatch performance.
Specialised techniques have been devised for this problem
that allow extremely fast dispatch of long runs using SIMD
and bitwise parallelism \cite{cameron2008,li2017,langdale2019}.
However, including whitespace and comments that must be processed
by the lexer (but not necessarily emitted), non-parenthesis tokens
are on average approx. 4 bytes long in the 2025 SMT-LIB corpus,
so it is necessary to optimise for short tokens instead.
As we limit \textbf{blazing-smtlib} to safe Rust only,
we dispatch a byte of the run by using it as an index into a 256-entry table,
where a byte value masked with the run flag determines if the run is continuing.
We loop this byte dispatch until the run ends, unrolling the loop 4 times.
Our experiments showed that this is on average comparable performance-wise with less portable
SIMD designs and superior to portable bitwise-parallel techniques due to the short token lengths.
For long tokens (relevant especially to the theory of strings),
switching to SIMD and long-token techniques at a certain point would presumably
be a faster but less portable alternative.

The category tests are also performed using a 256-entry table indexed by the current byte.
If the token cannot be processed by the eager dispatch loop, the lexer flushes the
token batch to the consumer and gracefully falls back to state-machine processing.
This can occur e.g. due to consuming all bytes in the provided byte slice or an unexpected
byte at the end of the run that will make the state-machine path produce an error.

We were not limited by stable Rust without unsafe constructs except for the aforementioned
inability to use SIMD explicitly and one lifetime restriction: since a zero-copy batch references
the input bytes and the lexer loses control over them when it returns,
it must either reassign the batch references or drop the batch structure
and create a new batch in the next call.

\subsection{Parsing}

Token consumers are provided a cursor through which they can peek and consume the next token
or attempt to consume multiple tokens at the same time. As the token provider is not required to
produce multiple tokens per batch, the attempts must fall back gracefully, so they are only
used in the library for the most likely and time-critical paths when parsing terms.
The consumers are required to consume at least one token per call so that parsing terminates.

For fast parsing, the parser adapter constructs resources for standard parsers,
which contain a state slot for every standard category, as well as typed arenas
(dynamic arrays of types that are pushed on the stack during processing).
This means that the constructed instances of e.g. \texttt{SmtTermParser} are zero-sized types
with no information, sharing a state slot and dynamic arrays so that already allocated memory
can be reused as much as possible.

User parsers cannot use our arenas except through a standard parser,
but a custom user arena could be threaded in via the handler at the cost of losing genericity.
To avoid inadvertent corruption of parser state through resource reuse,
the contract requires continuing to call parsers once they have started processing:
for example, it is forbidden to construct two instances of \texttt{SmtTermParser} and call them
alternately.

Most SMT-LIB constructs are easy to parse but there are two difficult types:
\begin{itemize}
  \item \textbf{Bindings.} Binders \smt{let}, \smt{forall}, \smt{exists},
    and \smt{lambda} open a new semantic
    scope\footnote{The binder \smt{match} functions similarly but
    opens scopes at its match cases instead.},
    and it is natural to provide the
    name-term bindings in a single call such as \texttt{start\_let} that will be followed
    by \texttt{end\_let}. For this, the bindings must be remembered during processing
    (we currently do not perform zero-copy binding dispatch for simplicity).
    Similar processing applies to binder-like commands such as \smt{define-fun}.
  \item \textbf{Recursion.} No standard SMT-LIB construct can alternate between
    different constructs recursively, but sorts, terms, and general S-expressions
    may self-recurse. The main implementation concept is keeping a stack of unclosed values
    and a parser state for parsing the current value at the top.
\end{itemize}

\begin{figure}[t]
  \includegraphics{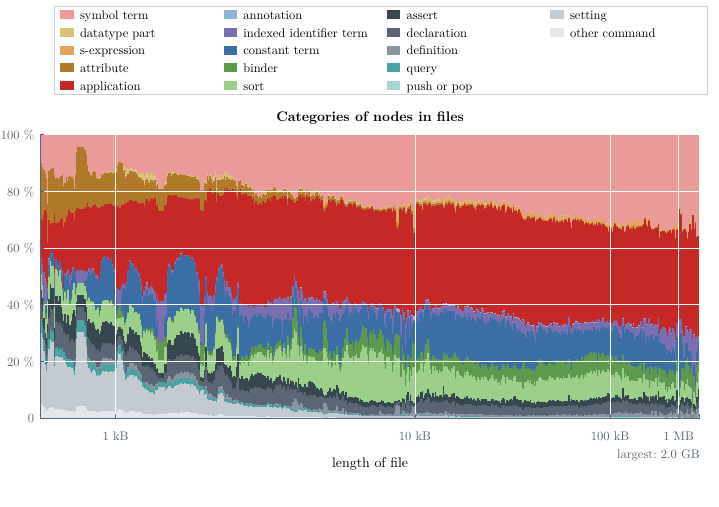}
  \caption{
    Shares of category and command nodes on 2025~SMT-LIB corpus files.
    Terms are dominant, especially as the file size increases.
    Symbol and constant terms can be dispatched to the handler immediately.
    Applications and indexed identifiers are common, but simple in nature.
    Binders are the most challenging because of their shape,
    need to remember the bindings, and uncommon occurrence
    that complicates branch prediction.
  }
  \label{fig:node-shares}
\end{figure}

The term parser is the most difficult part of the parser implementation
as it uses both types of constructs and terms (especially symbol terms and applications)
are the most common nodes in SMT-LIB files, as shown in Figure \ref{fig:node-shares}.
We do not use any techniques other than careful choices of data structures
and cursor attempts for a consistent but only small speed benefit.

\subsection{Switching Lexical Modes}

SMT-LIB allows a clean split into a lexing and a parsing pass,
with two notable exceptions:
\begin{itemize}
  \item \textbf{Exit command.} The \smt{(exit)} command forces the reader
    to immediately stop reading the SMT-LIB script.
  \item  \textbf{Version switch.}
    The version under which an SMT-LIB file should be parsed can be given by the
    command \smt{(set-info :smt-lib-version <version>)}. The legacy version 2.0 used a different escape format for string literals than the current versions from 2.5 onward.
    As the version is declared inside the script file, and it is read in a streaming fashion,
    the command switches not just the grammar and semantics,
    but the lexicon itself.

\end{itemize}

Parsing these constructs interacts with lexing, as errors and tokens produced
eagerly by the lexer may no longer be applicable after they are processed.
We resolve them carefully by allowing the consumer to request the new version
or discard mode through the token cursor, which will occur after the last consumed token.
The version switching from/to 2.0 in particular forces our lexer to be speculative:
after such a switch, it discards the tokens it already produced and starts lexing again
from the end of the last consumed token. This is possible since all tokens in the
current batch correspond to the bytes in the byte buffer provided
to the lexer as the input, which are still available to the lexer.

%% file: sections/05_lean.tex
\section{Lean Port}
\label{sec:lean}

We ported \textbf{blazing-smtlib} into Lean as \textbf{BSmtLib}.
Like the Rust version, it has no dependencies (except for Lean itself)
and contains no unsafe code. Additionally, it is panic-free:
all contract violations that result in a panic in Rust are
impossible by construction as the contract is formalised
and adherence to it must be proven. All proofs in the library
use only the three usual Lean axioms (classical choice,
propositional extensionality, and the quotient axiom),
and adherence to these properties is checked by an automatic
audit tool that also runs \textbf{leanchecker} and \textbf{nanoda} \cite{nanoda}
for additional sanity checking.

Porting \textbf{blazing-smtlib} to Lean was not trivial.
Despite the fact that both languages
are influenced by the language ML,
their main intended use is different: Rust is a systems programming language
with focus on speed and avoidance of memory manipulation bugs,
while Lean is a language with focus on theorem proving.
Their type systems thus excel at different tasks over a classic typed language.
Rust allows precisely controlling memory ownership and borrowing,
while Lean allows precisely controlling contract requirements and guarantees
as well as proving propositions about behaviour.

During porting, we discovered that most Rust constructs we use
map into Lean fairly well.
However, since Lean does not offer precise control
over memory and mutation, it is difficult to write highly performant code
as the decisions on memory layout, allocations,
and mutable assignments are made by the compiler.
As an unfortunate allocation on a hot path can decrease
the performance drastically when performed for each byte or token,
fine-tuned optimisations were much more crucial than in the case of Rust.

Even with proper inlining directives, we found our ported code to be
slower than desirable\footnote{We show the measurements of ported and optimised versions in Figure \ref{fig:eval-summary}.},
so we wanted to optimise it through rewriting without impacting correctness.
Fortunately, this is a speciality of Lean:
it is possible to write theorems about expected behaviour
of e.g. a lexer and prove that both the ported and the optimised version fulfill the contract,
avoiding bugs as long as the theorems are stated correctly and fully describe the expected behaviour.

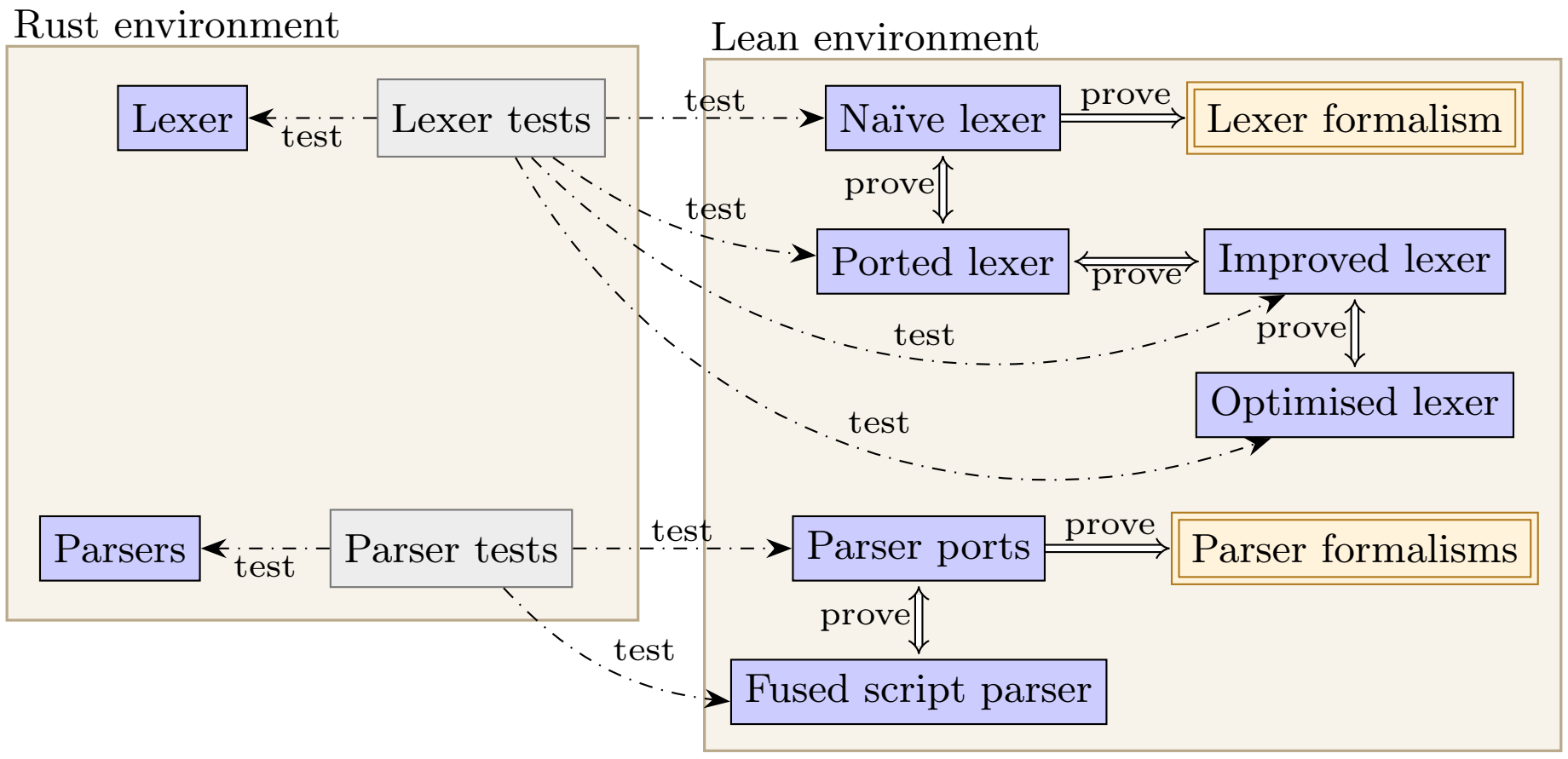
\begin{figure}[tb]
  \centering
  \begin{tikzpicture}[node distance=0.6cm and 1.0cm]

    \node[component] (rust-lexer)  {Lexer};
    \node[test] (lexer-tests)  [right = of rust-lexer] {Lexer tests};

    \node[component] (lean-lexer-v)  [right = 1.7cm of lexer-tests] {Naïve lexer};
    \node[component] (lean-lexer-w)  [below = of lean-lexer-v] {Ported lexer};
    \node[formalisation] (lean-lexer-spec)  [right = of lean-lexer-v] {Lexer formalism};
    \node[component] (lean-lexer-x1)  [below = of lean-lexer-spec] {Improved lexer};
    \node[component] (lean-lexer-z)  [below = of lean-lexer-x1] {Optimised lexer};


    \node[formalisation] (lean-parser-spec)  [below = of lean-lexer-z] {Parser formalisms};
    \node[component] (lean-parser)  [left = of lean-parser-spec] {Parser ports};
    \node[test] (parser-tests)  [left = 1.7cm of lean-parser] {Parser tests};
    \node[component] (rust-parser) [left = of parser-tests] {Parsers};

    \node[component] (lean-parser-fused)  [below = of lean-parser] {Fused script parser};

    \draw[tested] (lexer-tests) -- node[rel] {test} (rust-lexer);
    \draw[tested] (lexer-tests) -- node[rel] {test} (lean-lexer-v);
    \draw[tested] (lexer-tests) to[bend right=15] node[rel] {test} (lean-lexer-w);
    \draw[tested] (lexer-tests) to[bend right=35] node[rel] {test} (lean-lexer-x1);
    \draw[tested] (lexer-tests) to[bend right=40] node[rel] {test} (lean-lexer-z);

    \draw[proved] (lean-lexer-v) -- node[rel] {prove} (lean-lexer-spec);
    \draw[equal] (lean-lexer-w) -- node[rel] {prove} (lean-lexer-v);
    \draw[equal] (lean-lexer-x1) -- node[rel] {prove} (lean-lexer-w);
    \draw[equal] (lean-lexer-z) -- node[rel] {prove} (lean-lexer-x1);


    \draw[tested] (parser-tests) -- node[rel] {test} (rust-parser);
    \draw[tested] (parser-tests) -- node[rel] {test} (lean-parser);
    \draw[tested] (parser-tests) to[bend right=20] node[rel] {test} (lean-parser-fused);

    \draw[proved] (lean-parser) -- node[rel] {prove} (lean-parser-spec);
    \draw[equal] (lean-parser-fused) -- node[rel] {prove} (lean-parser);

    \begin{scope}[on background layer]
      \node[env, fit=(rust-lexer) (lexer-tests) (parser-tests) (rust-parser)] (rust-env) {};
      \node[env label] at (rust-env.north west) {Rust environment};
      \node[env, fit=(lean-lexer-spec) (lean-lexer-v) (lean-lexer-w) (lean-lexer-x1) (lean-lexer-z)
      (lean-parser-spec) (lean-parser) (lean-parser-fused), inner sep=2mm] (lean-env) {};
      \node[env label] at (lean-env.north west) {Lean environment};

    \end{scope}
  \end{tikzpicture}
  \caption{How the library components are checked against each other for consistency.
    Tests may not catch all bugs, and proving that the behaviour of lexers and parsers
  follows their formalisms adds an additional assurance.}
  \label{fig:consistency}
\end{figure}

In addition to running Lean lexers and parsers
through the Rust tests we use to pin the behaviour of \textbf{blazing-smtlib},
we formalised our reading of the SMT-LIB syntax as described
in the specification documents in Lean,
and proved that the lexers, parsers, and reader correspond to the expected behaviour
with handlers that reject non-standard commands.
We visualise how we detect inconsistencies between our lexers, parsers, and their formalisms
in Figure~\ref{fig:consistency}.
Every inconsistency detected while working on optimising
is then a problem that triggers an investigation,
minimising the potential for uncaught bugs.

The lexer proofs were done successively through these links:
\begin{enumerate}
  \item The lexer formalism.
  \item A naïve lexer that steps through the input slices byte by byte, performing no batching.
  \item A faithful Rust port, using runs, batches and eager zero-copy dispatch,
    throwing away batches when each lexer call ends (discussed in Subsection~\ref{subsec:lexing}).
  \item An improved lexer that avoids throwing away batches and just clears their length
    (unlike Rust, this poses no problems in Lean).
  \item The optimised lexer that turns the \texttt{Nat} (big integer) types
    inside the hot path into \texttt{USize} (machine-sized type) where possible in correspondence to Rust \texttt{usize}.
    As the natural way to work with numbers (including indexes) in Lean is by using big integers,
    \texttt{USize} handling is tricky and requires careful changes.
    This lexer is exposed to the users as it is the fastest.
\end{enumerate}

The parser proofs were done successively through these links:
\begin{enumerate}
  \item The formalisms of all parsers.
  \item A faithful Lean port of all parsers. We did not port token-consuming attempts for simplicity.
  \item A fused script parser. It walks the buffer of tokens itself instead of using
    peek/next on the batch cursor for maximum performance, and reduces the number of functions
    by essentially manual inlining.
\end{enumerate}

While our formalisation and proof effort found no faults with reading SMT-LIB files,
we found two places in our combinators where they called an inner parser with an empty cursor,
in violation of the parser contract. It also prompted us to rethink our mechanisms of lexer-mode changes
and locations where a handler should be called (e.g. always on the closing parenthesis for command ends),
ultimately leading to a cleaner and more consistent API.

\subsubsection{API and Naming.}
Some differences are forced by the languages' import and naming discipline.
In Rust, we keep every public item prefixed \texttt{Smt}, as library
types such as \texttt{Symbol} or \texttt{Term} could collide with the user types
for these concepts, cluttering autocompletion.
In Lean, it is more usual to refer to types prefixed by the namespace,
so this problem disappears.
However, there are many places where the namespace name must be written.
As \texttt{BlazingSmtlib} would be cumbersome to type, we opted for \texttt{BSmtLib} in the Lean port.

%% file: sections/06_evaluation.tex
\section{Evaluation}
\label{sec:evaluation}

We have evaluated the performance of our library on 9,201 files
from the 2025 SMT-LIB corpus against comparable tools:
publicly available SMT-LIB parsers and solvers.
Our evaluation compared lexer performance, full-parse performance (checking for syntactic but not semantic errors),
and semantic performance (checking for both syntactic and semantic errors).

\subsubsection{Tool search.} We performed a search of SMT-LIB parsers and tools
and rejected those that are not production-viable
due to commonly encountered stack overflows, timeouts under generous conditions, etc.
We also excluded \textbf{aws-smt-ir}, as its lexing and parsing results are
almost exactly the same as the much more popular \textbf{smt2parser},
which it is based on.

\subsubsection{Benchmark selection.} For benchmarks, we applied
the SMT-COMP 2026 rule~4 \cite{smt-comp-rules}
to reduce the weight of overrepresented benchmarks in the corpus and
speed up the measurement.
After selecting one file per family,
we have drawn the others via sorting them by the SHA-256 hash of their file paths.
This resulted in selection of 78,010 files.
We excluded 57 files over 200 MB from evaluation
due memory and time limits
and 68,752 files under 200 kB
as unsuitable for evaluation as their startup time
significantly impacts the data\footnote{
  We show the data for our decision
  in Figure \ref{fig:small-files} in Appendix \ref{app:measurements}
}.
As such, we used 5,629 non-incremental and 3,572 incremental files for the plots and evaluation.

\subsubsection{Lexing and parsing.} For parsers, we wrote probes that apply the most similar conditions to all parsers:
the file is loaded into memory, the in-process measurement is started, its bytes are fed to the parser
(as a batch if possible), and the measurement is ended once processing is done.
A handler that counts commands is used, and we print the success/error result and count afterwards
to make sure the parser checks the syntax.
We also measured lexing performance for parsers where the lexers
are separable from parsers, using the same probe but only running the lexer.
For a sanity check, we ran the probes for our tools and the \textbf{Dolmen} probe in a hash-building mode,
and they agreed on all corpus files except two where \smt{(exit)(exit)} appears.
In parsing mode, \textbf{Dolmen} continues after the first \smt{(exit)} while our parsing stops.

\subsubsection{Well-formedness.} As we wanted to show the real-life performance a user may expect when using our library,
we wrote a toy well-formedness checker that accepts or rejects in accordance with SMT-LIB
semantics; this requires keeping bindings, operations, etc. in memory similarly to a solver.
After a preliminary comparison with strict \textbf{Dolmen} type-checker,
we adjusted the checker behaviour on some constructs so it agrees with \textbf{Dolmen} on acceptance
of files in the corpus.
We then measured against \textbf{Dolmen}, our solver \textbf{Roole}, and other well-known solvers
on non-incremental files stripped of commands such as \texttt{(check-sat)}.
While we tried to make the measurement as fair as possible,
it represents an unusual mode of operation for solvers:
they may use techniques such as term sharing that
decrease command-processing performance in favour of faster solving.
The measurements do not suggest anything about solving performance.

\subsubsection{Measurement setup.}
We ran our measurements on an x86\_64 AMD Ryzen 9950X3D machine
running under Windows WSL2 with a 78 GB memory limit.
We measured in parallel on 16 cores,
with a 60 second timeout per run.
The first run decided acceptance and was timed,
then repeated up to 9 more times if running under a 100 millisecond budget.
For the final results, we chose the fastest run.
We only used files accepted by the tools (not errors or timeouts)
for visualisation and computation of geometric mean.

\begin{figure}
  \includegraphics{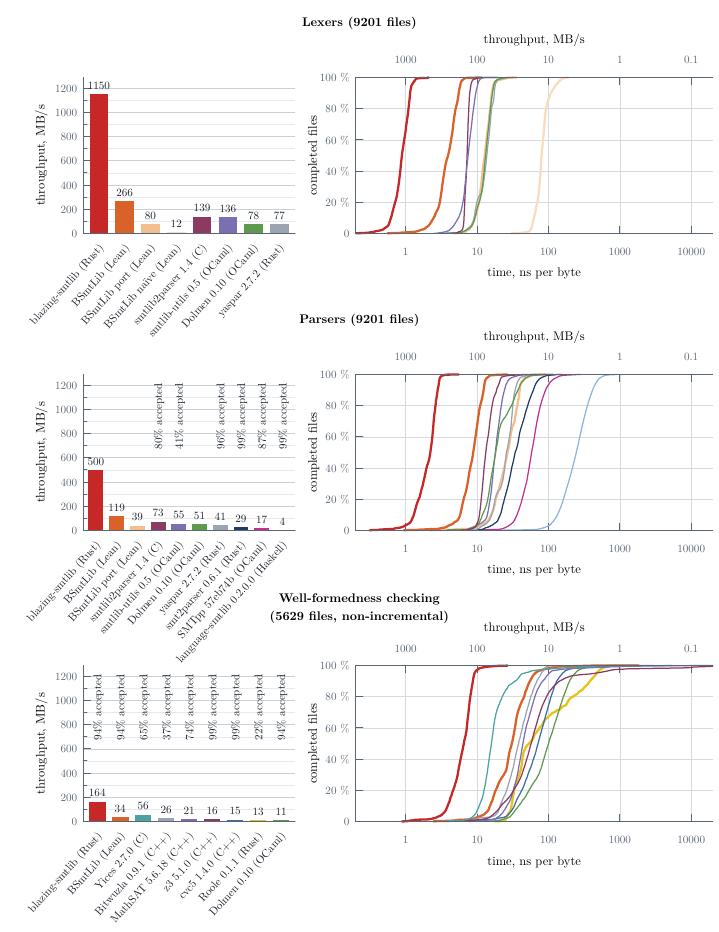}
  \caption{
    Measurements of our tools against other lexers, parsers, and executable tools
    that can be used to check well-formedness of SMT-LIB scripts.
    The left column shows the geometric-mean throughput. The right column shows the
    percentage of files that are processed with at least the given throughput,
    with time-axis differences corresponding to a speedup/slowdown by a ratio.
    Most tools have a highly consistent shape; some solvers perform
    noticeably slower processing on certain files,
    which lowers their mean but may be desirable to speed up solving.
  }
  \label{fig:eval-summary}
\end{figure}

\begin{figure}
  \includegraphics{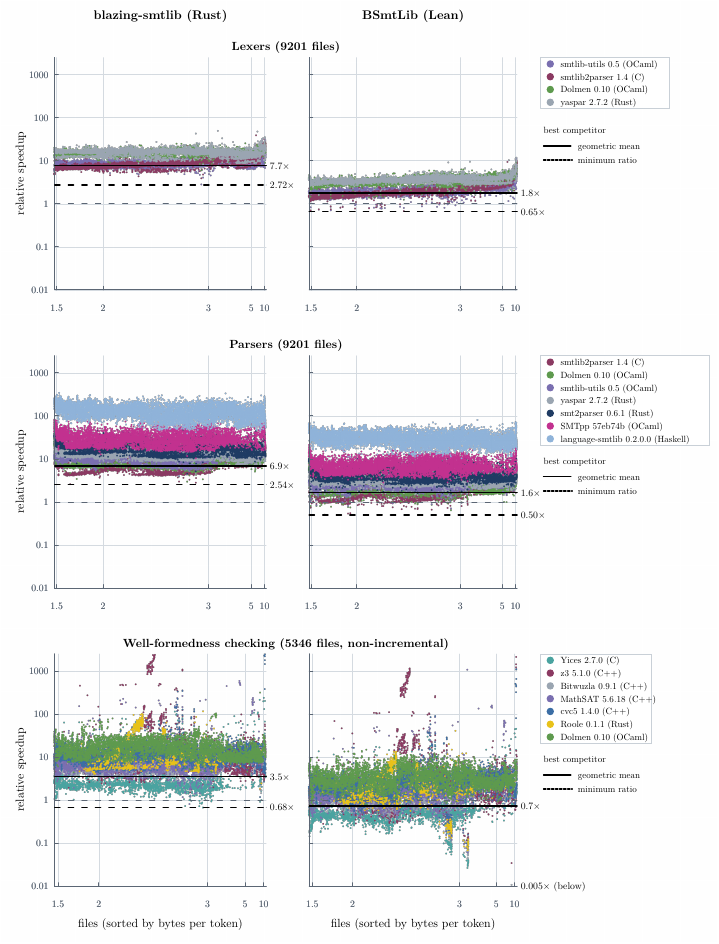}
  \caption{
    Ratio of time taken by \textbf{blazing-smtlib}, \textbf{BSmtLib},
    and the well-formedness checkers
    against other lexers, parsers, and executable tools.
    Each file/solver combination is shown separately.
    The files are sorted by the average number of bytes per token.
    Speedups tend to be consistent, longer tokens exhibiting
    slight trends in some comparisons.
    On files with very long tokens, some tools become
    slower relatively to ours due to their higher per-byte work.
  }
  \label{fig:eval-ratios}
\end{figure}

\subsection{Measurement Results}

Our lexer and parsers have accepted all of the measured files,
with reported token/command counts equal to those reported
for other lexers/parsers that accepted the file.
Our well-formedness checker matched \textbf{Dolmen} in acceptance
where \textbf{Dolmen} reached a verdict.

We show our cleaned measurement results in Figure \ref{fig:eval-summary},
and the speedup ratios of \textbf{blazing-smtlib} and \textbf{BSmtLib} in Figure \ref{fig:eval-ratios}.
\textbf{blazing-smtlib} is at least 2.5\texttimes{} faster than the best competitor
on every file, and more than 6\texttimes{} on a geometric mean of files.
\textbf{BSmtLib} is more than 1.5\texttimes{} faster than the fastest competitor
on a geometric mean of files, although only 0.5\texttimes{} as fast on worst-case files.

In well-formedness checking,
our Rust well-formedness checker outperforms all other tools
on geometric mean, but loses to \textbf{Yices 2.7.0} on 10 files
and to \textbf{MathSAT 5.6.18} on 1.
The Lean well-formedness checker holds its ground to most competitors
on geometric mean, but \textbf{Yices} is faster, even 200\texttimes{}
faster in the worst case.
We investigated the \textbf{Yices} source code; its lexer and parser
are written very efficiently, allowing it to reach this level of performance despite
the additional burden of performing term sharing.

%% file: sections/07_conclusion.tex
\section{Conclusion \& Future Work}
\label{sec:conclusion}

We presented new parsing libraries for SMT-LIB~2,
\textbf{blazing-smtlib} written in Rust
and its port \textbf{BSmtLib} written in Lean.
We discussed their composable API which allows extension
to non-standard SMT-LIB commands,
the implementation in Rust with the major speed-related decisions,
its port with Lean-specific optimisations
(the potential for bugs minimised by testing and
correspondence to our formalised reading of SMT-LIB syntax),
and evaluated their performance.
On a representative sample of the 2025~SMT-LIB corpus,
\textbf{blazing-smtlib} was more than 6 times faster
than previous parsing libraries on the geometric mean,
and \textbf{BSmtLib} was faster or comparable to them.
Our libraries can ease the development of
SMT-LIB solvers and proof-checkers, metadata summarisers,
and even tools requiring extensions such as optimisation solvers
and SV-LIB verifiers.
Lean users can now also easily use the SMT-LIB corpus for tasks such as
optimising automated tactics.
In general, our libraries let
users care less about the SMT-LIB parsing and its performance
and focus on their core tasks.

In the next release of our tools \textbf{Roole} and \textbf{Roolean},
we plan to switch the used parsers to \textbf{blazing-smtlib} and \textbf{BSmtLib},
respectively.
Since the parser is a part of the trusted codebase
in \textbf{Roolean}~\cite{onderka2026twin},
adding a formalisation of SMT-LIB semantics built on
our formalised reading of SMT-LIB syntax would allow even greater trust
in the proof-checker result.

\begin{credits}
  \subsubsection{\ackname}
  This work was partially funded by a gift from Amazon.

  \subsubsection{\discintname}
  The authors declare that there is no other competing interest besides the
  acknowledged funding and the support through their affiliations.

  \subsubsection{AI Declaration.}
  We used Claude Code (mainly the models Fable 5.1 and Opus 5.5)
  for development of the library and the artefacts supporting this paper,
  including those used to generate figures.
  This paper is completely human-written,
  with Claude Code used only in a helper role,
  especially for finding references and checking consistency.
  AI-generated code is reviewed with scrutiny corresponding
  to its importance.

  In the Rust version of \textbf{blazing-smtlib},
  AI is used only for experiments and carefully reviewed changes
  where the semantic shape is clear but manual implementation cumbersome,
  with the API shape of the library decided by the maintainer
  (first author).

  The Lean version, a port from Rust with Lean-specific changes,
  is written and managed by AI under human guidance, including proofs.
  Our formalised reading of SMT-LIB syntax was generated by AI,
  reviewed by the maintainer and iterated upon.

  The tests, written in Rust and used to test both versions of the library,
  are managed by AI under human guidance. The 2025 SMT-LIB corpus is used as
  an additional test.
\end{credits}

\subsubsection{Data Availability Statement.}
\textbf{blazing-smtlib} and \textbf{BSmtLib} are available in a single GitHub
repository \cite{blazing}. Additionally, \textbf{blazing-smtlib} is published on
the persistent Rust package repository \href{https://crates.io/}{crates.io}
\cite{blazing-published}. The artefact to this paper that contains tools for
reproduction of measurements and figures is available at
\url{https://doi.org/10.5281/zenodo.23362806} \cite{artefact}.

%% file: sections/a_corpus.tex
\clearpage
\section{Additional Corpus Data for Lexer Choices}
\label{app:corpus}

\begin{figure}[!ht]
  \includegraphics{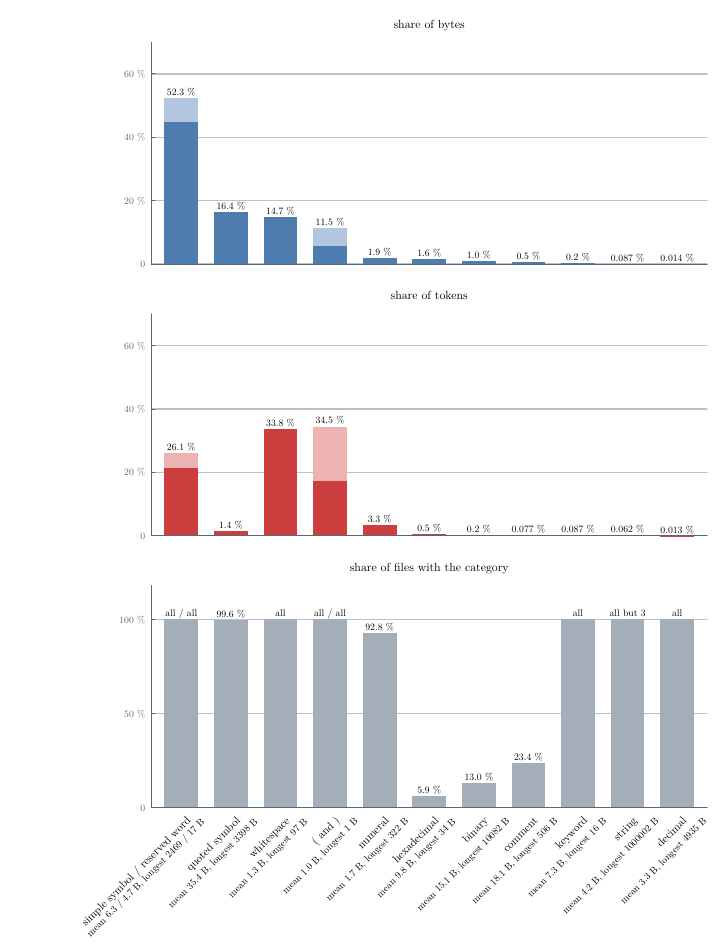}
  \caption{
    Share of bytes and tokens in the whole 2025 SMT-LIB corpus
    when considering layout (whitespace and comments).
    Symbols, whitespace, and parentheses dominate
    both in bytes and token counts, with over half of bytes being symbols.
  }
  \label{fig:categories}
\end{figure}

\begin{figure}[p]
  \includegraphics{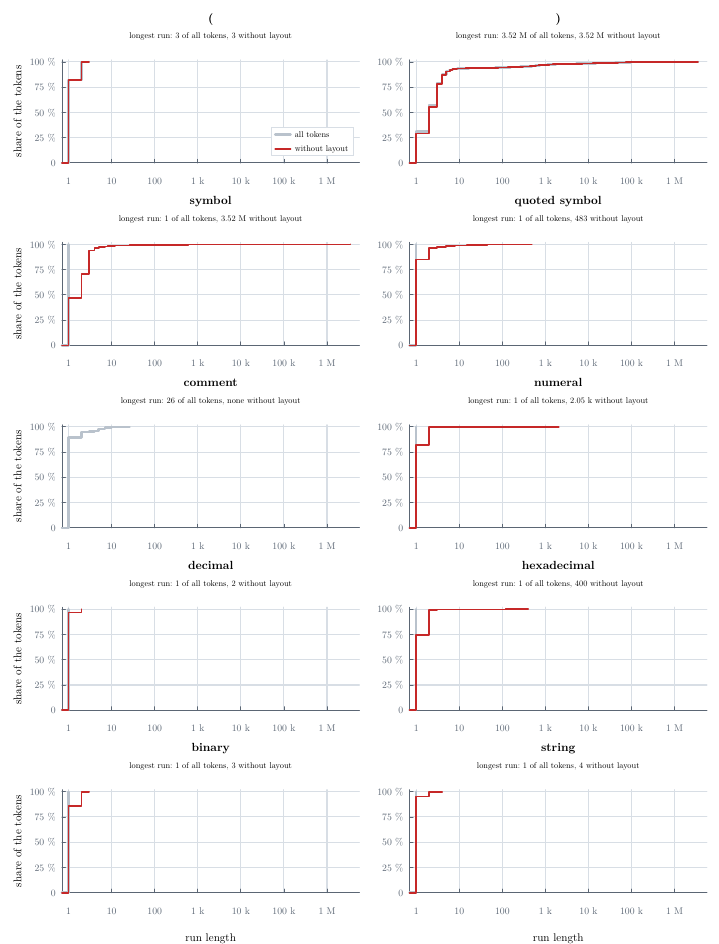}
  \caption{
    Token runs in the whole 2025 SMT-LIB corpus,
    when excluding layout (whitespace and comments)
    and considering it either end the run or continue.
    There are almost no runs except for
    heavy runs of the closing parenthesis
    in some files and runs interrupted by whitespace
    for symbols.
    This is the reason for the preference of the closing parenthesis
    in Subsection \ref{subsec:lexing}.
  }
  \label{fig:token-runs}
\end{figure}

\begin{figure}[p]
  \includegraphics{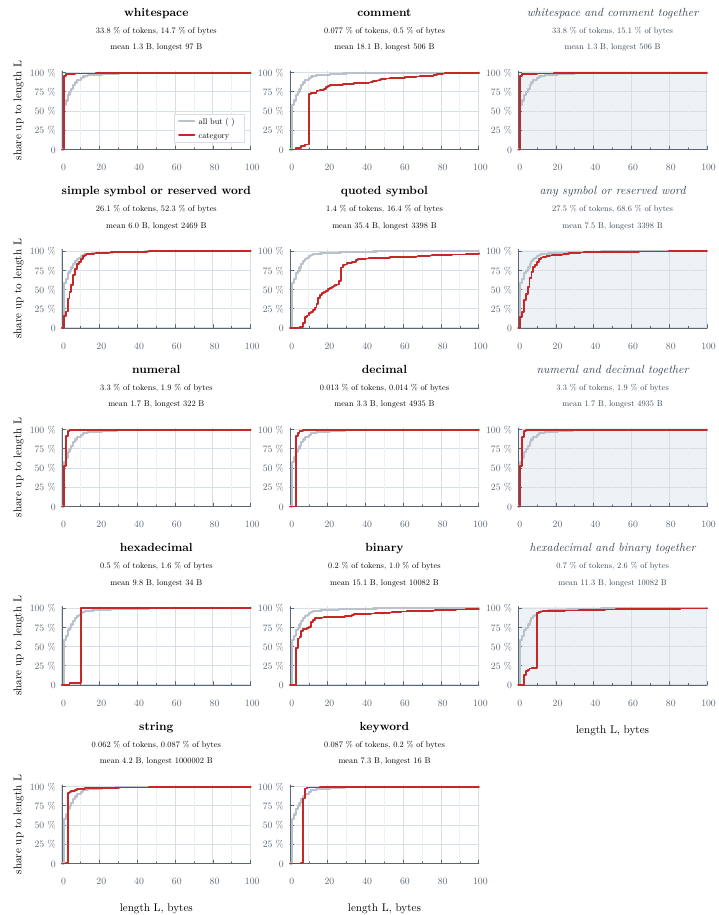}
  \caption{
    Token lengths in the whole 2025 SMT-LIB corpus,
    including layout (whitespace and comments).
    Overall, the token lengths are usually short, under 10 bytes,
    usually making long-run techniques unnecessary.
    The major exception to this are comments and quoted symbols,
    which are however not represented in the corpus enough to
    matter much. The jump in comment size is due to the comments
    \smt{; success} (10 bytes including its semicolon
    and line terminator).
    The longest string in the corpus is a generated adversarial
    input. Its content without the enclosing double quotes is 1 MB.
  }
  \label{fig:token-lengths}
\end{figure}

%% file: sections/b_measurements.tex
\clearpage
\section{Additional Measurement Data}
\label{app:measurements}

\begin{figure}[!ht]
  \includegraphics{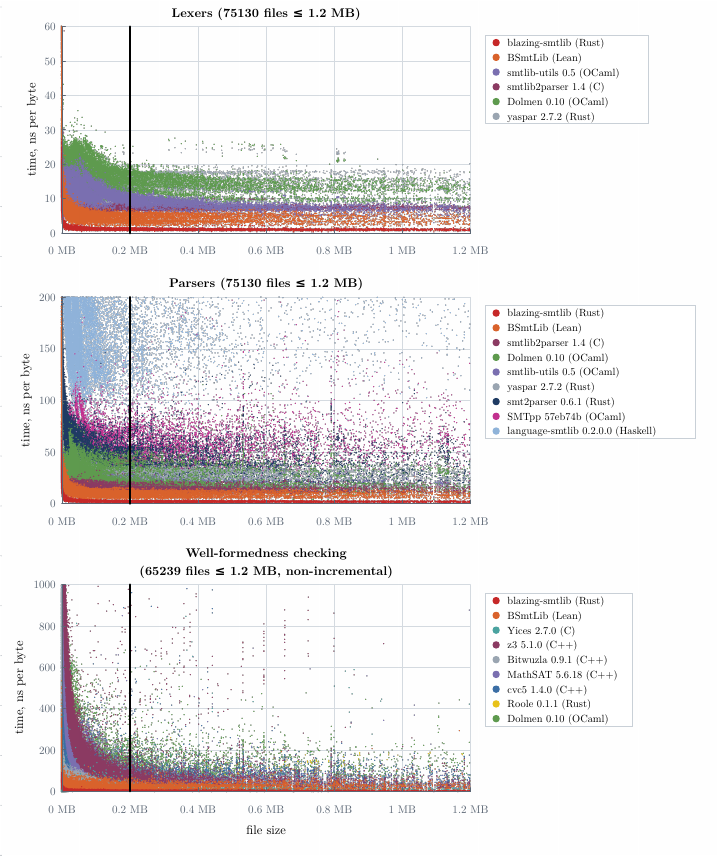}
  \caption{
    The reason for our decision to remove files under 200 kB from evaluation.
    The plot is linear over ns/byte and file size
    (i.e. lesser is better), zoomed in
    to show the startup behaviour.
    The startup impact is significant for small files even when performing
    in-probe measurement, as we do on the lexers and parsers.
    We consider 200 kB to be a reasonable choice that mostly sanitises the startup effect
    and keeps enough benchmarks for measurement.
  }
  \label{fig:small-files}
\end{figure}

\begin{figure}[p]
  \includegraphics{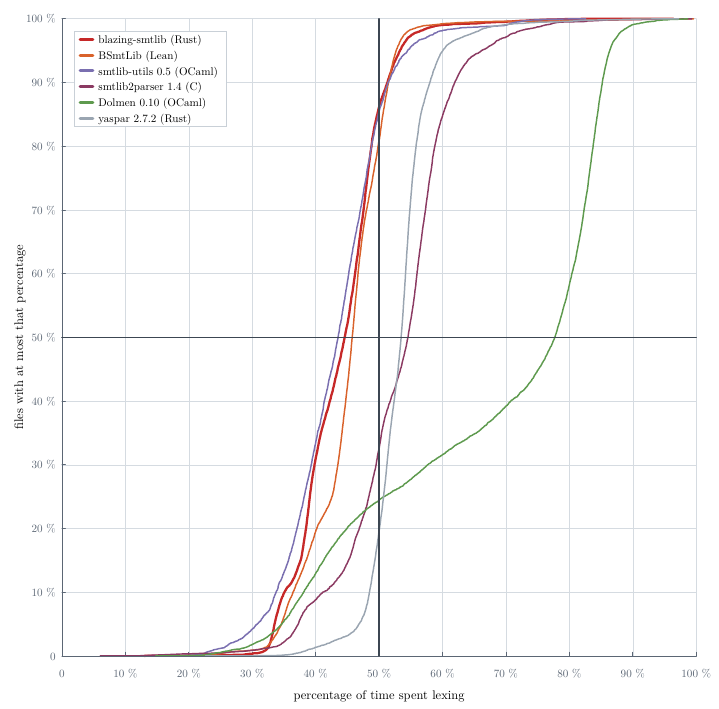}
  \caption{
    How much work is spent lexing in comparison to the full
    lexer-parser pipeline.
    Most parser libraries spend around 50\%, suggesting the lexing side and the parsing side
    are balanced and it is not easy to improve the combined performance
    by only speeding up one of them. \textbf{Dolmen} is a notable exception.
  }
  \label{fig:lex-parse}
\end{figure}